\documentclass[aip,reprint]{revtex4-1}
\usepackage[utf8]{inputenc}

\usepackage{graphicx}
\usepackage{amsmath}
\usepackage{gensymb}
\usepackage{textgreek}
\usepackage{physics}
\graphicspath{ {Figures/} }

\draft 

\begin{document}


\title{A Femtosecond Magnetic Circular Dichroism Spectrometer} 



\author{Jake Sutcliffe}
\affiliation{EaStCHEM School of Chemistry, University of Edinburgh, Edinburgh, UK}
\author{J. Olof Johansson}
\email[]{olof.johansson@ed.ac.uk}
\affiliation{EaStCHEM School of Chemistry, University of Edinburgh, Edinburgh, UK}


\date{\today}

%

\begin{abstract}
    We describe the development of a broadband magneto-optical spectrometer with femtosecond temporal resolution.  The absorption spectrometer is based on a white-light supercontinuum (\textit{ca}. $320 - 750$~nm) using shot-to-shot temporal and spectral referencing at 1~kHz. Static and transient absorption spectra using circularly polarised light are collected in a magnetic field. The difference spectra with respect to the external field direction give the static and transient magneto-optical Faraday rotation (magnetic optical rotary dispersion) and ellipticity (magnetic circular dichroism) spectra. An achromatic quarter-wave plate is used and the impact of the deviation from ideal retardance on the spectra is discussed. Results from solution-based and thin-film samples are used to demonstrate the performance and wide applicability of the instrument. The sensitivities for the static and time-resolved data were found to be 5 and 0.4~mdeg, respectively. The method presents a simple way to measure magneto-optical spectra using a transient absorption spectrometer and an electromagnet. 
\end{abstract}

\maketitle

\section{Introduction}
    The photo-induced dynamics in
    magnetic materials on ultrafast timescales have received much interest over the past two decades, \cite{Kirilyuk2013LaserAlloys,Bigot2013UltrafastNanostructures,Walowski2016PerspectiveSpintronics} mainly driven by the search for ever faster data storage technologies. Such research has been made possible through femtosecond (fs) magneto-optics, where the time-dependent change in polarisation state (rotation or ellipticity) of a probe pulse in the presence of a magnetised material is recorded. This is possible because magnetised materials change the rotation and/or ellipticity of transmitted and reflected light through the Faraday and Kerr effects, respectively. Application of this method led to the the first observation of demagnetisation of thin films of ferromagnetic Ni on the 100~fs timescale.\cite{Beaurepaire1996UltrafastNickel}  
    
    Transient changes in absorption and polarisation can vary greatly depending on the probe wavelength used. Much more information about the system can be obtained if multiple wavelengths are probed. Khorsand \textit{et al.} used wavelength tuneable probe light, which enabled them to achieve element specificity in a TbFe alloy so that they could  measure the dynamics in each magnetic sublattice.\cite{Khorsand2013Element-SpecificLight} By probing either side of the band gap of the ferromagnetic semiconductor EuO, Formisano \textit{et al.} could disentangle the dynamics of short- and long-range magnetic interactions.\cite{Formisano2020FemtosecondEuO} Recent developments in ultrafast magnetism have resulted in photo-magnetic recording in dielectric materials.\cite{Stupakiewicz2017UltrafastMedium} This has the distinct advantage of faster sequential switching rates due to the non-thermal mechanism. \cite{Stupakiewicz2017UltrafastMedium} These experiments make use of specific d-d crystal field transitions, which in turn changes the magneto-crystalline anisotropy.  \cite{Stupakiewicz2017UltrafastMedium,Stupakiewicz2019SelectionGarnet, Mikhaylovskiy2020ResonantOxides, Baierl2016TerahertzReversal} A spectroscopic approach might shed more light on the problem. In addition to dielectrics, other new materials in ultrafast magnetism are now being pursued, such as magneto-plasmonic crystals, \cite{Dyakov2019Widecrystals,Frolov2020Darkcrystals} molecule-based magnets, \cite{Johansson2016DirectlyTime-resolution} and two-dimensional ferromagnets.\cite{Fang2018LargeTe6,DeSiena2020} However, much of the research devoted to ultrafast magneto-optics to date has only considered quasi-monochromatic light because of the lack of spectral features in reflectance and transmission spectra of metallic samples. 
    
    For optical wavelengths, an optical parametric amplifier (OPA) can be used as a probe to record the signal one wavelength at a time by measuring the change in ellipticity using bridged photodiodes.\cite{Mertens2020WideFields,Khorsand2013Element-SpecificLight} Likewise, it is possible to select a narrow band of a white-light supercontinuum and record one wavelength at a time.\cite{Johansson2016DirectlyTime-resolution} However, measuring the whole spectrum at once is much more efficient, which can be achieved using a white-light supercontinuum spectrally dispersed onto a CCD array in combination with a set of polarisation optics.\cite{Bigot2004FemtosecondOptics} Due to the wavelength dependence of the sample and the optical components, one has to be careful when measuring the spectrally resolved polarisation change. This lead Bigot \textit{et al.} to record the magneto-optical spectra as a function of the angles of the polarisation optics, which allowed them to extract the ellipticity spectra.\cite{Bigot2004FemtosecondOptics} However, these extra measurements add to the total measurement time.   
    
  
    There is an alternative way of recovering the magneto-optical signals by measuring the absorption of left- and right-handed circularly polarised light in a magnetic field. This leads to magnetic circular dichroism (MCD) spectra, which are equivalent to wavelength-dependent changes in ellipticity. \cite{Hiramatsu2015Communication:Spectroscopy,Mertens2020WideFields} Such spectra are a valuable analytical technique for investigating many substances from inorganic complexes to large proteins, and can reveal much about the electronic structure of compounds. MCD has also shown to play an important role in fs all-optical switching in ferrimagnetic GdFeCo and granular L1$_0$ FePt ferromagnets. \cite{Ellis2016Alldichroism, Khorsand2012MCDRecording} To obtain an MCD spectrum, the difference in absorbance of right- and left-handed circularly polarised light by a sample in a magnetic field is measured. Typically, MCD spectra are measured by rapidly changing the handedness of the light incident of the sample through the use of a photoelastic modulator. \cite{Mollenauer1969StressMCD} Using a photoelastic modulator in time-resolved measurements still limits one to measuring a single wavelength at a time. \cite{Oppermann2019UltrafastUltraviolet}
    Instead of reversing the handiness of the probe beam, one can reverse the direction of the external magnetic field and recover the MCD spectrum.\cite{Boschini2015AMagneto-optics} This approach was used by Yao \textit{et al.} to record X-ray MCD spectra using high harmonic generation. \cite{Yao2020ARange} This method is also suitable when using a white-light supercontinuum. However, care has to be taken because white-light continua typically show strong temporal and spectral noise correlations.\cite{Bradler2014TemporalSpectroscopy} Shot-noise limited spectra can be achieved with transient absorption (TA) spectroscopy based on supercontinuum generation by making use of shot-to-shot referencing by recording pump on/off and simultaneously measuring the spectrum of a reference beam. \cite{Lang2018PhotometricsArt} Our TA spectrometer makes use of these methods and we here show how the inclusion of a magnetic field can be used to record time-resolved MCD (TRMCD) spectra.
    
    In this paper, we describe a broadband TRMCD spectrometer in transmission (Faraday) geometry. Unlike measurements in reflection geometry (magneto-optical Kerr effect), the transmission geometry also allows us to investigate liquid samples. Samples are excited with either 800 or 400~nm pump pulses (or a non-collinear OPA if a wider wavelength range is required) and probed with a white-light supercontinuum. After conversion into circularly polarised light, the light passes through the sample and is dispersed and detected by a CCD array. Spectra are recorded for opposing field directions and the difference between the two gives the MCD. Shot-to-shot spectral fluctuations in the intensity of the white-light are accounted for by the inclusion of a second CCD array to measure a reference signal. With this, submillidegree sensitivities can be achieved for time-resolved measurements. The addition of a polariser allows the same setup to also measure both static and time-resolved Faraday rotation spectra. Jones calculus is used to justify this approach, and also to quantify the effect of a nearly-achromatic quarter-wave plate. Finally, static spectra for solid and solution-based samples and time-resolved spectra for a ferromagnetic thin film are measured to show the wide applicability of this setup.

\section{Methodology}
\begin{figure*}
        \centering
        \includegraphics[width=0.8\textwidth]{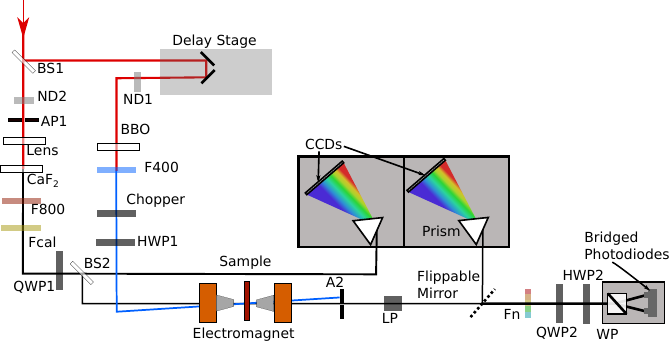}
        \caption{Simplified schematic of the setup, where the labels stand for: neutral density filter (ND), aperture (AP), quarter-wave plate (QWP), half-wave plate (HWP), linear polariser (LP) and Wollaston prism (WP). Several filters are also used: an 800~nm shortpass filter (F800), a 400~nm bandpass filter (F400), a calibration filter (Fcal) and several interchangable bandpass filters (Fn). QWP1 is removed when measuring Faraday rotation and LP is typically removed when measuring MCD.}
        \label{fig:setup}
    \end{figure*}

    The transient absorption spectrometer used in this work is described in Ref. \onlinecite{Liedy2020VibrationalPhotoexcitation} and is based on the one described in Ref. \onlinecite{Megerle2009CaF2WLG}. An amplified Ti:Sapphire laser system produces 120~fs pulses of 800~nm light at a repetition rate of 1~kHz (top-left of Fig. \ref{fig:setup}). A beam splitter (BS1) is used to separate out the pump and probe pulses. 
    
    The pump beam is directed into a motorised delay stage. After a time delay is introduced by the stage, a neutral density filter (ND1) is used to adjust the pulse energy of the pump. The pump can be frequency-doubled by a beta barium borate (BBO) crystal and the remaining 800~nm light filtered out. A chopper blocks every second pulse, before the pulse passes through a half-wave plate (HWP1), which is used to control the polarisation direction of the pump. The pump beam is focused onto the sample using a 50~cm spherical mirror. An aperture (AP2) centred on the probe beam is used to block the residual pump beam after the sample and a linear polariser (LP) can be inserted to further reduce the amount of scattered pump light from the sample.
    
    The probe beam passes through a neutral-density filter (ND2), an aperture (AP2), and a lens which optimise the beam for generation of a white-light supercontinuum in a CaF\textsubscript{2} plate. The remaining 800~nm light is removed by a 750~nm shortpass filter, resulting in a spectrum ranging from roughly $320 - 750$~nm. However, the intensity decreases greatly below 400~nm, and the light is unstable above approximately 650~nm. Circularly polarised light is generated by an achromatic quarter-wave plate (QWP1: Thorlabs AQWP05M-600). The low transmittance of the QWP below 400~nm removes most of the already low-intensity UV part of supercontinuum, reducing the range to around $400 - 650$~nm. However, QWP1 can be substituted for one more suitable for the UV so that wavelengths down to 300~nm can be investigated, albeit at the cost of a much reduced range at longer wavelengths. To generate a reference signal, a beam splitter (BS2) diverts half of the beam towards the detection stage. The probe beam is focused using a 50~cm focal length spherical mirror through the poles of the electromagnet onto the sample, where it overlaps with the pump beam. The angle between the pump and probe beams is $1.2\degree$. By aligning the probe beam with holes drilled through the poles of the magnet, both the probe and magnetic field are normal to the sample.  
     
    After the sample, the probe beam is then directed through a prism and focused onto a CCD array, identical to that which measures the spectrum of the reference beam. These two CCD arrays each consist of 512 pixels, which measure the light incident on them each millisecond, and have a resolution of approximately 0.7~nm per pixel. The signals from the 512 pixels in each CCD array are exported to a computer for further calibration and post-processing. A spectral filter (Fcal: Thorlabs FGB67M) with a rich and detailed absorption spectrum is inserted into the beam and used for wavelength calibration. This also ensures that the same pixels on the reference and probe CCDs correspond to the same wavelengths. A MATLAB script is used to fit a polynomial to correct for the chirp in the white-light. 
     
    The sample is placed between the poles of a 0.4~T electromagnet (GMW 3470) with 10~mm diameter axial holes in the poles. To achieve a 0.4~T field, a current of 5~A is applied by a unipolar power supply (Aim-TTi CPX400SP). In order to reverse the field, the supply is turned off, the current direction reversed and then turned back on. A cryostat can also be placed in the sample position to carry out low temperature measurements. 

    A pair of bridged photodiodes (Thorlabs PDB210A/M) used in conjuction with a lock-in amplifier (Zurich Instruments MFLI) provides another method of measuring femtosecond magneto-optics. Using a bandpass filter to select a narrow (c.a. 10~nm) wavelength range, a mirror is flipped to allow the filtered beam into a wollaston prism (WP) which splits the beam into perpendicular polarisation components that are detected with two photodiodes. The lock-in amplifier with a reference frequency set to the repetition rate of the laser then detects the difference in signal between the two photodiodes. A half-wave plate (HWP2) directly before the photodiodes is used to balance the photodiodes by minimising the difference signal. Because of this, a non-zero difference signal corresponds to a change in rotation of the incoming light. A change in ellipticity can be measured using a quarter-wave plate (QWP2) in front of HWP2. For ultrafast magneto-optics, the reference frequency is halved to match that of the pump pulses, so that only changes to the polarisation state due to the pump pulse are recorded. These photodiodes are used with a 460~nm bandpass filter to ensure that QWP1 is at the correct angle to circularly polarise the probe beam.
    
    For static magneto-optics, a continuous wave 450~nm diode laser (ThorLabs CPS450) can also be used. The laser is modulated by a chopper and attenuated using an ND filter. After this, it is directed along the path of the probe beam into the photodiodes and then detected by the lock-in amplifier at the frequency of the chopper. Despite the single-wavelength limitation, the diode laser is much more intense and stable than the white-light continuum. Therefore, we often use the diode laser for magnetic-field sweeps in static magneto-optic measurements.
    
    A number of samples are used to show that the results obtained through this setup are as expected. To test the setup's ability to measure rotation of light, the Verdet constant of a 1.1~mm thick soda-lime glass (Thermo Scientific BS7011) was measured, as this is well-studied over a range of wavelengths. The MCD and absorption spectra of CoCl\textsubscript{2}$\cdot$6H\textsubscript{2}O both change depending on whether the compound is dissolved in water or concentrated HCl. Solutions of the compound at a concentration of 0.125~M were made up with distilled water and $\sim$37\% HCl, and then decanted into a cuvette with a path length of 1~cm. MCD spectra were also measured for two ferromagnetic samples: a Ni film and a CoPt bilayer with in- and out-of-plane anisotropies, respectively. The Ni film was grown through electron-beam physical vapor deposition on soda-lime glass and is estimated to be around 20~nm thick. The CoPt bilayer was grown on glass with the form Ta(4~nm)/Pt(3.5~nm)/Co(1.04~nm)/Ta(5~nm), where the Ta capping layers protect the film.

\section{Static Measurements}
    \subsection{Faraday Rotation}
        The static Faraday rotation spectra measurements are first discussed before detailing how the MCD spectra are recorded.  The rotation of light caused by the sample in a magnetic field can be measured using a polariser after the sample (LP in Fig. \ref{fig:setup}). QWP1 is not used in this configuration. The linearly polarised white-light supercontinuum passes through the sample, the LP, and is subsequently dispersed onto the CCD to measure the spectrum. Jones calculus is used to quantify the effect of the magnetic field and the polariser angle on the measured spectrum. Light reaching the sample is horizontally polarised and described by the vector $\ket{p} = \begin{pmatrix} 1\\ 0 \end{pmatrix}$. It is assumed that the effect of the sample on the light itself is given by
        \begin{equation} \label{eq:sample}
        \mathbf{W} = 
            \begin{pmatrix}
            1 & \Theta \\
            -\Theta & 1
            \end{pmatrix},
        \end{equation}
        where $\Theta=\theta - i\eta$. Here $\theta$ and $\eta$ denote the rotation and ellipticity induced by the sample, respectively. However, $\theta$ and $\eta$ only correspond to the actual change in polarisation state when $|\Theta|<<1$ rad ($\sim 57$~deg). This condition is satisfied for all the samples and field strengths discussed in this paper ($<750$ mdeg). It is assumed that $\Theta$ is proportional to the applied field and so reversal of the applied field changes the signs of the off-diagonal terms in $\mathbf{W}$. 
        
        After transmission through the sample, the light passes through LP, which is described by
        \begin{equation} \label{eq:polariser}
            \mathbf{P} = 
            \begin{pmatrix}
                \cos^2 \beta & \cos \beta \sin \beta\\
                \cos \beta \sin \beta & \sin^2 \beta
            \end{pmatrix},
        \end{equation}
        where $\beta$ is the angle that the polariser's axis of transmission makes to the horizontal plane. The light reaching the CCDs is denoted $\ket{p'} = \mathbf{PW}\ket{p}$ and the measured intensity at the CCDs, $S$, is given by
        \begin{equation} \label{eq:intensity}
            S = \braket{p'}{p'}.  
        \end{equation}
        
        In the setup described in Fig. \ref{fig:setup}, two CCDs record the intensity of the probe and reference beams simultaneously. The transmittance, $T$, of the sample is given by the ratio of the intensity of transmitted light to the light's incoming intensity and is converted to absorbance through $A = -\log_{10} T$. This conversion neglects reflected or scattered light and so the absorbance only corresponds to absorbed light when there is negligible reflection. Furthermore, the probe and reference beams travel different paths into the CCDs, and so have slightly different spectra. Only the ratio of $T$ before and after the magnetisation reversal is relevant, so slight changes in spectra and non-negligible reflectance can be ignored.
        
        The probe beam passes through the sample and its intensity in the presence of a positive or negative magnetic field, $S^\pm$, is given by Eq. (\ref{eq:intensity}). Although the reference beam is unaffected by the sample, it is still subject to fluctuations in the supercontinuum generation so the reference intensities corresponding to $S^\pm$ are denoted $R^\pm$. The change in absorbance between positive and negative fields, $\Delta A$, can therefore be expressed as
        \begin{equation} \label{eq:mcdfromexp}
            \Delta A = A^+ - A^- = -\log_{10}\left(\frac{S^+ R^-}{R^+ S^-}\right).
        \end{equation}
        Since the ratio $R^-/R^+$ only accounts for fluctuations in the intensity of the light, it is assumed that $R^-/R^+ = 1$, which simplifies the subsequent calculations. Substituting Eqs. (\ref{eq:sample}) -- (\ref{eq:intensity}) into Eq. (\ref{eq:mcdfromexp}), it is found that
        \begin{equation}
            \Delta A = \log_{10} \left(\frac{\cos^2\beta + \sin^2\beta|\Theta|^2 - 2\theta \sin 2\beta}{\cos^2\beta + \sin^2\beta|\Theta|^2 + 2\theta \sin 2\beta}\right),
        \end{equation}
        where $\Delta A = 0$ if the polariser is parallel or perpendicular to the incoming light. However, when the polariser is intermediate at $\beta = \pi/4$, then $\Delta A \approx \left(4/\ln 10\right) \theta$ provided that $|\Theta|<<1$. This therefore provides a means through which we can measure Faraday rotation or, equivalently, the magnetic optical rotatory dispersion (MORD) spectrum. 
        
        To measure a static spectrum, $S^+$ and $R^+$ are recorded over a large number of laser pulses (typically 3000, which takes three seconds per acquisition given the 1~kHz repetition rate) and then $S^+/R^+$ is averaged. When the field is reversed, $S^-/R^-$ is obtained in the same manner so that Eq. (\ref{eq:mcdfromexp}) can be used to get the MORD spectrum. This is repeated as necessary, and we find that the sensitivity of static MORD (and MCD) is typically limited to around $5 - 10$~mdeg, although additional averaging can reduce this further. This noise level is larger than that of typical static MCD spectrometers due to the fluctuations inherent in the generation of a femtosecond white-light supercontinuum.
        
        \begin{figure}
            \centering
            \includegraphics[width=0.45\textwidth]{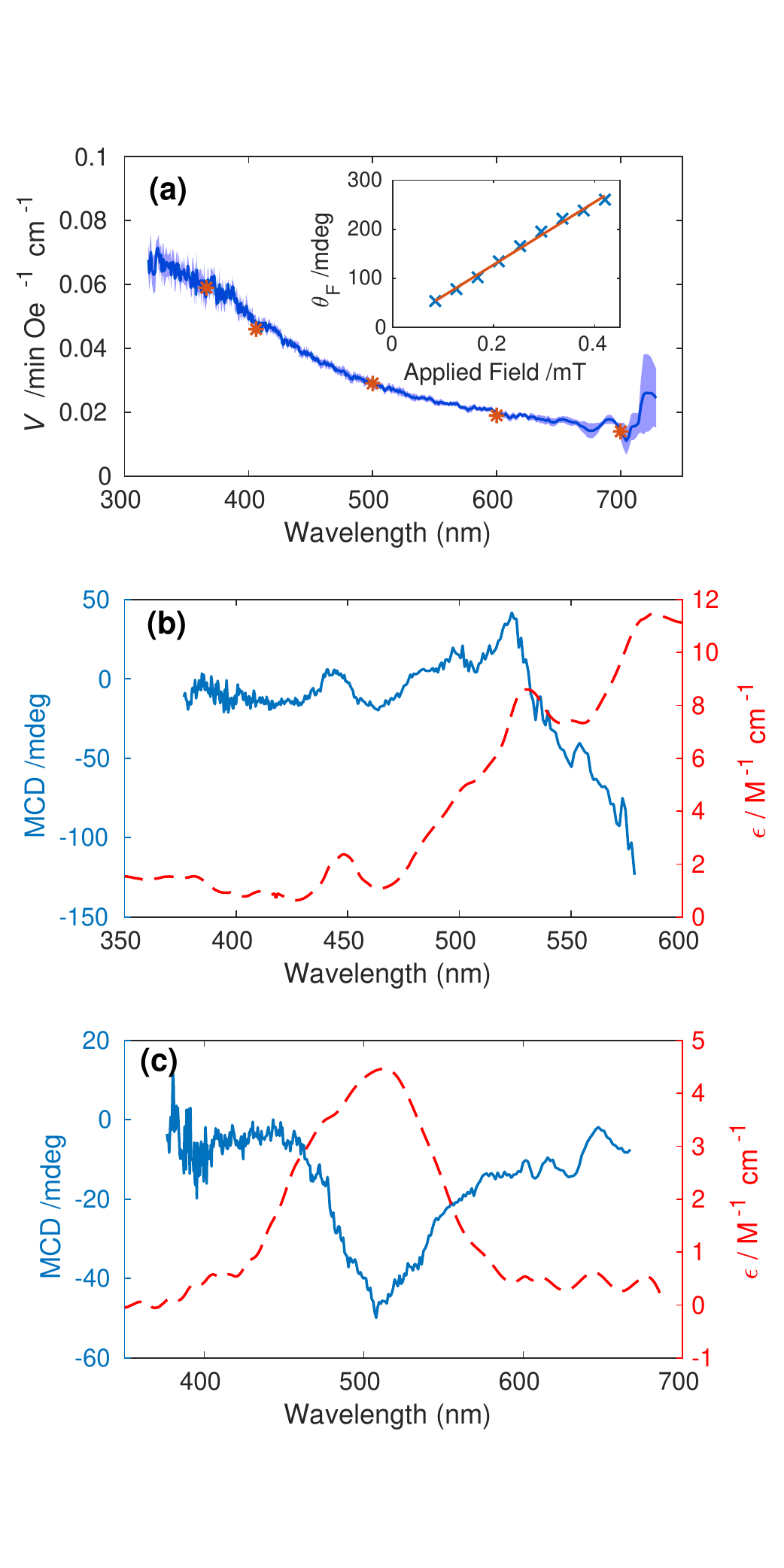}
            \caption{(a) Verdet constant, $V$, of soda lime glass, where the blue shaded area is the 68\% confidence interval and literature values \cite{Robinson1964The} are shown as red asterisks. Inset shows a typical linear fit for rotation, $\theta_F$, at 450~nm against applied field strength, used to calculate the Verdet constant. MCD (solid lines) and molar attenuation, $\epsilon$, (dashed lines) spectra of CoCl\textsubscript{2} in (b) water and (c) concentrated HCl. There is a 10~mdeg offset to the MCD spectrum of the acidic solution which has arisen from the removal of the solvent spectrum, but this has no impact on the overall shape of the spectrum. }
            \label{fig:cocl2andglass}
        \end{figure}
        
        To test the above theoretical description, the Verdet constant, $V$, of a 1~mm thick piece of soda-lime glass was measured. The Verdet constant is related to the Faraday rotation by $\theta_\textrm{F} = VBl$, where $B$ is the magnetic field strength and $l$ is the thickness of the material. MORD spectra of glass were measured for different applied field strengths and a linear fit to the data was carried out at each wavelength to determine $V$. The resulting spectrum is shown by the blue line in Fig. \ref{fig:cocl2andglass} (a) and the 68\% confidence intervals of the fit are shown by the shaded area above and below the line. Literature values \cite{Robinson1964The} are shown at several wavelengths marked by the red asterisks, and all lie within, or very close to, the error on the measured values. 
    
    \subsection{Faraday Ellipticity}
        If LP is removed and a quarter-wave plate (QWP1) inserted before the sample, ellipticity rather than rotation can be measured. Light circularly-polarised by QWP1 passes through the sample in the magnetic field and is then dispersed onto the CCD. In the same procedure as with the rotation measurements described above, the change in absorbance between positive and negative applied fields gives the ellipticity. To ensure that a wide range of wavelengths can be measured, QWP1 must be achromatic with a constant retardance $\rho = \pi/2$ across the range. No wave plate is truly achromatic, but if its retardance is known then its effect on the ellipticity can be calculated. The effect of the quarter-wave plate is therefore quantified as 
        \begin{equation} \label{eq:qwp}
            \mathbf{Q} = e^{i\rho/2}
            \begin{pmatrix}
                \cos^2 \alpha +e^{i\rho}\sin^2 \alpha & (1-e^{i\rho})\sin\alpha\cos\alpha\\
                (1-e^{i\rho})\sin\alpha\cos\alpha & \sin^2 \alpha +e^{i\rho}\cos^2 \alpha
            \end{pmatrix},
        \end{equation}
        where $\alpha$ is angle between the fast and horizontal axes. The light reaching the CCD is denoted $\ket{p'} = \mathbf{QW}\ket{p}$. From substituting this into Eq. (\ref{eq:intensity}) and simplifying, the intensity of the light in the presence of positive and negative fields is given by
        \begin{equation} 
            S^\pm = 1+|\Theta|^2 \pm 2\eta \sin 2\alpha \sin \rho.
        \end{equation}
        Since the change in rotation and ellipticity is small, $|\Theta|<<1$~rad, the total change in absorbance can be approximated from Eq. (\ref{eq:mcdfromexp}) as
        \begin{equation} \label{eq:dAapprox}
            \Delta A \approx -\frac{4}{\ln10} \eta \sin 2\alpha \sin \rho.
        \end{equation}
        Therefore, $\Delta A$ is proportional to the ellipticity. This proportionality is more clear if $\alpha = \pi/4$ and $\rho = \pi/2$. Converting $\eta$ to degrees and assumming a perfect wave plate, gives the result that
        \begin{equation} \label{eq:elipfromda}
            \eta\mbox{[degrees]} \approx 33 \Delta A \mbox{[OD]}.
        \end{equation} 
        The MCD can also be measured by directly measuring the change in absorbance of left- and right-handed circularly polarised light in a constant magnetic field. Analysis of such measurements \cite{PracticalMCD} also results in Eq. (\ref{eq:elipfromda}), provided that $\eta << 1$ rad, showing the equivalence between reversing the field and reversing the handedness of the light.
        
        If the wave plate is imperfect, then the effect of the wavelength-dependent retardance can be removed from the measured MCD spectrum by dividing by $\sin \rho$. However, the retardance of QWP1, as recorded by the manufacturer, varies at most by 20\% from its ideal value of $\pi/2$ rad. This therefore only reduces the MCD spectrum by 5\% at most, so retardance of the wave plate has a minimal impact on the measurements. 
        
        MCD spectra of CoCl\textsubscript{2}$\cdot$6H\textsubscript{2}O were measured to compare to MCD spectra presented in the literature because it is a well-studied molecule. To extract only the signal from the Co(II) complexes, MCD and absorbance spectra of the pure solvents were measured and subtracted from the MCD and absorbance spectra of the aqueous and acidic solutions. The resulting MCD spectra are plotted as solid blue lines in Fig. \ref{fig:cocl2andglass} (b) and (c) for the aqueous and acidic solutions, respectively, and absorbance as dashed lines. Due to the large absorbance of the acidic solution of wavelengths greater than 550~nm, the MCD in this range in unusable. Despite this, the shapes of both absorbance and MCD spectra for both solutions match well with those reported in the literature, with clear peaks at 445, 500 and 525~nm in the acidic and 508~nm in the aqueous MCD spectra, which all lie within $<$5~nm of the literature values. \cite{Coleman1972MagneticAnhydrase,Schooley1965MagneticReport} There is little consensus in the literature as to the absolute magnitude of the MCD spectra themselves, but the spectra shown here are of the same order of magnitude. 
    
    \subsection{Ferromagnetic Thin Films}
\begin{figure*}
            \centering
            \includegraphics[width=\textwidth]{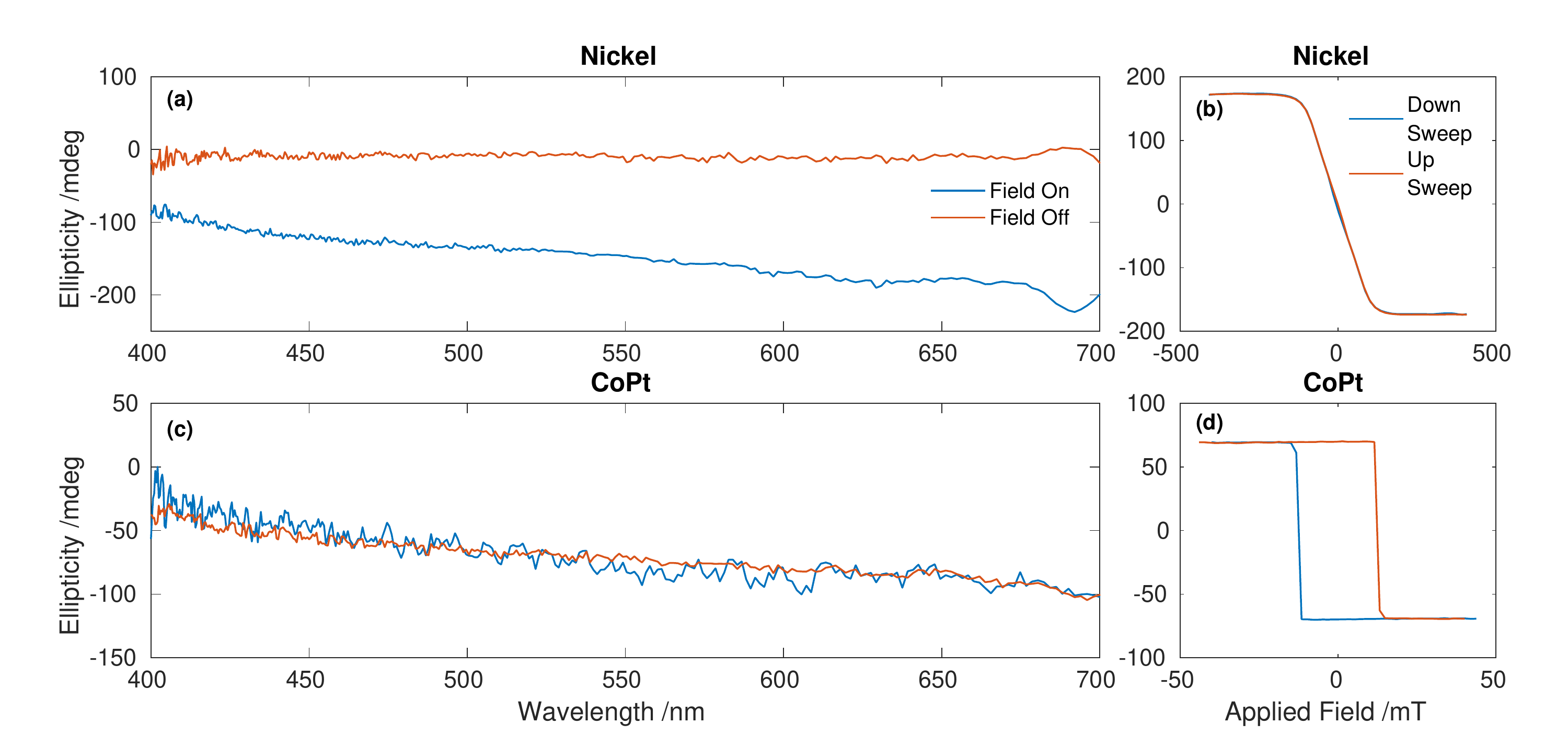}
            \caption{Static MCD spectra of thin films of (a) Ni and (c) CoPt bilayer with and without an external applied magnetic field. (b,d) Hysteresis loops for the same samples measured at 450~nm using a laser diode and bridged photodetectors. Both samples are ferromagnetic, but only the CoPt has a remanent magnetisation normal to the sample plane, which is reflected in the square hysteresis loop and the lack of the difference between MCD spectra with the field on and off. The intensity of the supercontinuum below 400~nm is very weak and is unstable above 650~nm, which reduces the signal-to-noise in these spectral regions.}
            \label{fig:fieldonoff}
        \end{figure*}

        MCD spectra were also measured for two ferromagnetic samples: a Ni film and a CoPt bilayer with in- and out-of-plane anisotropies, respectively. These spectra are presented in Fig. \ref{fig:fieldonoff} (a) and (c). Both spectra show an increase in the MCD signal towards longer wavelengths, but the MCD of the substrate also contributes somewhat to this shape. However, the spectra are different when measured with no applied field. Faraday rotation and ellipticity depend only on the magnetic field parallel to the path of the light and so only the out-of-plane magnetisation component is detected in the measurement. The difference in anisotropy between the samples is illustrated in their hysteresis loops in Fig. \ref{fig:fieldonoff} (b) and (d), which were measured using the diode laser and bridged photodiodes. As can be seen, there is no hysteresis present for the nickel film but a clear remanence in the CoPt bilayer is observed. Typically, the magnetic field is applied constantly while MCD spectra are collected, and spectra measured in this manner are shown as the blue lines in Fig. \ref{fig:fieldonoff} (a) and (c). However, given the ferromagnetic nature of these samples, there will still be a remanent magnetisation within the sample even if no external field is applied. To investigate this, MCD spectra were measured by applying fields only to reverse the magnetisation direction within the sample, and then turning them off again while the spectra were collected. The resultant spectra are plotted in Fig. \ref{fig:fieldonoff}, where the difference between the two samples is clear. Since the Ni film has no remanence normal to the sample plane, MCD is only observed if the external signal is applied. In contrast, whether there is an applied field or not has little impact on the CoPt spectrum, reflecting the out-of-plane magnetisation present in the sample. Both samples have relatively low transmittance but the CoPt transmits less, and this results in a higher level of noise in the CoPt data in Fig. \ref{fig:fieldonoff} (c). The MCD spectra also include the MCD of the substrate, hence the discrepancy between the saturation values of the hysteresis loops and the values of the MCD at 450~nm (this signal was subtracted from the measurement with the bridged photodiodes). 

\section{Time-Resolved Measurements}
    To record time-resolved changes in absorbance, the difference between the steady-state absorbance spectrum, $A_0$, and the spectrum at time delay $t$ following a pump pulse, $A(t)$, is measured. By blocking every second pump pulse, the difference between adjacent shots gives the transient absorbance. Reference spectra are measured simultaneously alongside every laser shot. The TRMCD in degrees, $\Delta A_{MCD} (t)$, is therefore given by
    \begin{eqnarray} \label{eq:TrdAMCD}
        \Delta A_{MCD}(t) &&= \left[(A^+(t) - A^-(t)) - (A^+_0 - A^-_0) \right] \nonumber\\
        &&= -\frac{1}{2}\log_{10} \left(\frac{S^+(t)  R^-(t) R^+_0 S^-_0}{R^+(t) S^-(t) S^+_0 R^-_0} \right) \\
        &&\approx \frac{1}{33}\left(\eta(t)-\eta_0 \right) \nonumber
    \end{eqnarray}
    using the same notation as with the static MCD, Eq. (\ref{eq:mcdfromexp}). 
    Experimentally, a transient absorbance spectrum is measured for each field direction with circularly polarised light, and the difference in absorbance between the two is found. As with the static spectra, the inclusion of the reference spectra in Eq. (\ref{eq:TrdAMCD}) greatly improves the sensitivity of the time-resolved spectra. Typically, each delay point is averaged over 2000 consecutive pulses, resulting in a sensitivity of 1~mdeg, due to the shot-to-shot and spectral referencing. This value can be decreased to 0.4~mdeg when averaging over a narrow spectral range, as discussed below. This sensitivity is of the same order of magnitude as the sensitivities of other broadband (magnetic) circular dichroism spectrometers. \cite{Trifonov2010BroadbandControl, Hiramatsu2015Communication:Spectroscopy,Oppermann2019UltrafastUltraviolet} Instruments with higher sensitivities use much longer integration times \cite{Stadnytskyi2018NearSpectrometer} (several minutes) or use an OPA to only probe one wavelength at a time. \cite{Mertens2020WideFields}
    
    If a sample does induce transient changes to the polarisation state of the transmitted light, then $S^+ \neq S^-$. Therefore, to recover the TA spectrum, $\Delta A_{TA} (t)$, the average of the spectra measured in opposing fields is taken. This modifies Eq. (\ref{eq:TrdAMCD}) to:
    \begin{equation} \label{eq:TAfromMCD}
        \Delta A_{TA}(t) = \left[(A^+(t) + A^-(t)) - (A^+_0 + A^-_0)\right]/2.
    \end{equation}
    This means that both the TA and TRMCD are measured simultaneously.

    Time-resolved changes in the Faraday rotation spectrum can also be measured with Eq. (\ref{eq:TrdAMCD}) once the setup is rearranged into the same geometry used for the static rotation measurements. Since the polariser cuts out half of the light, rotation measurements are slightly less sensitive than ellipticity measurements.
    
    For pump-probe measurements, pump light scattered by the sample often dominates at wavelengths close to the pump wavelength in the resulting spectrum. In regular TA spectra, if the pump and probe are both linearly polarised with perpendicular planes of polarisation, a linear polariser can be placed after the electromagnet and adjusted so that all the pump light is blocked. Of course, care has to be taken to avoid anisotropy affects, but there are ways around this. \cite{Maly2017polarisationSpectroscopy} We have found that, with the right angles, it is possible to use the polariser to remove scattered light even when measuring MCD spectra. In order to investigate how this impacts the ellipticity measurements, the Jones vector description detailed above is used again. 
    The light reaching the CCDs is now given by $\ket{p'} = \mathbf{PWQ}\ket{p}$. Substituting this into Eqs. (\ref{eq:intensity}) -- (\ref{eq:mcdfromexp}), under the assumption that $|\Theta| << 1$, it is found that
    \begin{eqnarray} \label{eq:qwpandpolariser}
        \Delta A &&\approx \frac{4}{\ln 10}\frac{\theta \sin 2\beta \cos \rho +\eta \sin \rho}{1 + \cos \rho \cos 2 \beta}\\
        && = \frac{4}{\ln 10} \eta \frac{\sin \rho}{1+\cos \rho} \; \; \; \mbox{for } \beta = 0 \nonumber
    \end{eqnarray}
    When the polariser is parallel or perpendicular ($\beta = 0$ or $\pi/2$) to the incoming light, then $\Delta A$ is proportional to $\eta$ but is much more dependent on the retardance of the wave plate than Eq. (\ref{eq:dAapprox}). If the polariser angle is intermediate between these, then $\Delta A$ depends not only on $\eta$ but also $\theta$. We have set $\beta=0$ in the experiments. A polariser can therefore be used to remove pump scatter, although the results would have to be corrected for the imperfect retardance of the quarter-wave plate. Furthermore, the intensity of the light reaching the CCD is approximately halved by the polariser, which decreases the sensitivity of the spectrometer compared to time-resolved measurements that exclude the polariser. 
 
\begin{figure*}
        \centering
        \includegraphics[width=\textwidth]{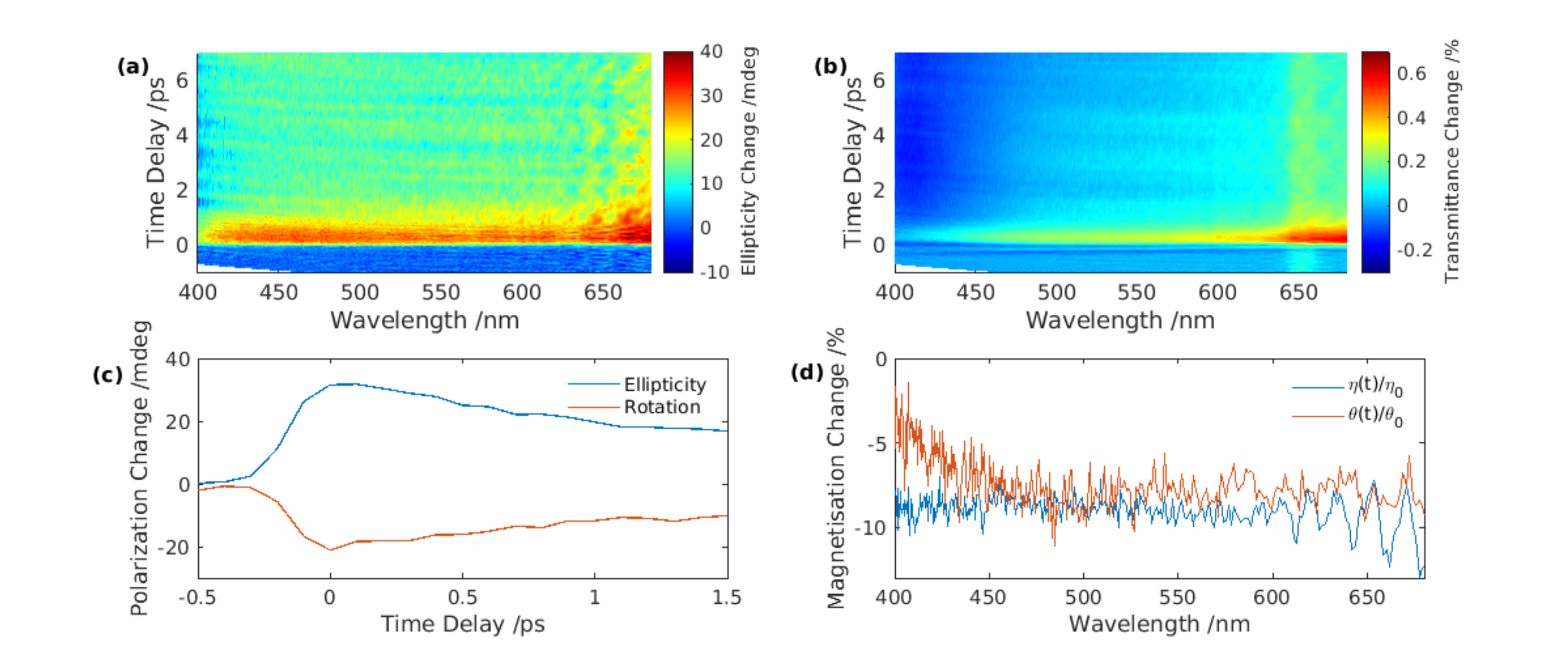}
        \caption{Demagnetisation of a Ni film exposed to an 800~nm, 5~mJ cm\textsuperscript{-2} pump pulse. (a) Time-resolved MCD and (b) Transient transmittance spectra, where the signal below 400~nm is reduced due to the small intensity of the supercontinuum at shorter wavelengths. The pump scatter can also be seen at wavelengths other than 800~nm and is characterised by a signal at negative time delays. (c) Transient ellipticity and rotation at 550~nm, averaged over 5~nm. (d) Magnetisation change at $t = 1.5$~ps after the pump pulse, calculated from both ellipticity and rotation. Dividing by the much noisier static spectra increases the noise level but both show roughly the same demgnetisation across the spectrum.}
        \label{fig:NickelTRMCD}
    \end{figure*}

   Time-resolved measurements were carried out at room-temperature on the aforementioned Ni film.  A pump wavelength of 800~nm and fluence 5~mJcm$^{-2}$ was used, and a  polariser used to remove the pump scatter. An external magnetic field of 0.4~T field was applied. The TRMCD and TA were calculated from the raw data using Eqs. (\ref{eq:TrdAMCD}) and (\ref{eq:TAfromMCD}), respectively, and then corrected for the impact of the polariser with Eq.\ref{eq:qwpandpolariser}. The resulting contour plots are shown in Fig. \ref{fig:NickelTRMCD} (a) and (b). Since the film has non-negligible reflectance, the TA is converted into percentage change in transmittance (by multiplying by a factor of $-100\ln 10$) to better represent the sample properties. The pump clearly induces a change in both the transmittance and MCD, which has clear spectral dependence and persists well beyond the duration of the pump pulse itself. There is a rapid demagnetisation with a rise time of 0.3~ps, which then decays with an exponential decay of 2~ps, after which the spectrum stabilises to a plateau. Such dynamics can be explained phenomenologically with the three-temperature model.\cite{Beaurepaire1996UltrafastNickel} 
    
    TRMCD and TRMORD spectra were measured for the same sample and fluence, but a 750~nm shortpass filter (Thorlabs FESH0750) was used to remove scattered 800~nm pump light instead of a polariser. A 5~nm wide gaussian weighted average was applied to the spectra at 550~nm and the resulting transient changes in the ellipticity and rotation are plotted in Fig. \ref{fig:NickelTRMCD} (c). The noise level of the CCD array is 1~mdeg for single pixels but the weighted average reduces this to 0.4~mdeg. This is over an order of magnitude smaller than the sensitivity of the bridged photodiodes when used to measure the same demagnetisation. Measurements with CCDs and photodiodes take roughly the same amount of time, but the CCD arrays are more sensitive at low light levels than the bridged photodiodes. The CCDs also use the reference beam to account for shot-to-shot and spectral fluctuations. Of course, it is possible to bypass the lock-in amplifier and use a single-pulse detection scheme with the bridged photodiodes to significantly improve the signal-to-noise ratio. \cite{Werley2011PulsedCard} The main advantage of our method is the simultaneous detection of the full spectral range. 
    
    Dividing the TRMCD and TRMORD by their static counterparts gives the percentage magnetisation change. By subtracting the static MCD and MORD spectra of the glass substrate from those for the Ni sample, the spectrum of the film itself can be recovered. A peak demagnetisation of ca. 16\% is observed, which is of the expected order of magnitude for such films and fluences. \cite{Koopmans2009ExplainingDemagnetization, Roth2012TemperatureMechanism} The spectral dependence of the demagnetisation at 1.5~ps, calculated from both the MCD and MORD spectra, is plotted in Fig. \ref{fig:NickelTRMCD} (d). Since the static spectra have much higher levels of noise, the demagnetisation consequently has lower sensitivity than the TRMCD or TRMORD. The noise in the static spectra accounts for the decrease in magnitude below 450~nm and the oscillations above 600~nm. However, no other spectral dependence is observed and the demagnetisations calculated from the TRMCD and TRMORD are roughly the same. The magnetisation change should be independent of wavelength and method of measurement so this confirms that we are indeed measuring both static and transient changes in ellipticity and rotation. 
    

\section{Conclusion}
    In conclusion, we have developed a table-top fs magnetic circular dichroism spectrometer, which can measure both static and time-resolved spectra. The generation of a white-light supercontinuum with a CaF\textsubscript{2} plate and subsequent detection with a CCD array allow us to measure an entire spectrum at once. Through shot-to-shot referencing, the sensitivity of the setup is 5-10 mdeg for static spectra, but this reduces by an order of magnitude to 0.4~mdeg for time-resolved spectra. To demonstrate the wide applicability of the setup, the Verdet constant of glass, and the MCD and absorbance spectra of CoCl\textsubscript{2}$\cdot$6H\textsubscript{2}O solutions were measured and agreed well with literature. TRMCD, TRMORD and TA spectra for a Ni film were also measured and compared to static spectra to detemine the demagnetisation. These results clearly show the sensitivity afforded by this approach when using a white-light supercontinuum, as well as the increased efficiency of measuring a whole spectrum rather than a single wavelength. It is hoped that this sensitive and broadband spectrometer will help shed light on the magneto-optical properties and processes within magnetic materials, as well as paramagnetic compounds.

\begin{acknowledgments}
    The authors thank H. Lewis and O. Nerushev for growing the Ni film, and T. Moorsom and O. C\'espedes for growing the CoPt bilayer. The authors acknowledge financial support from EPSRC grants EP/S018824/1 and EP/V010573/1.
\end{acknowledgments}

\bibliography{references.bib}

\end{document}